\documentclass[a4paper,12pt]{article}
\usepackage{latexsym}
\usepackage{amssymb}
\usepackage{amsmath}
\usepackage{cite}
\usepackage{color}

\usepackage{microtype}

\numberwithin{equation}{section}
\allowdisplaybreaks

\makeatletter \@addtoreset{equation}{section} \makeatother

\def\ii{{\rm i}}

\usepackage{ytableau}

\ytableausetup{centertableaux, mathmode, smalltableaux}

\definecolor{blue-violet}{rgb}{0.54, 0.17, 0.89}
\definecolor{PineGreen}{cmyk}{0.92, 0, 0.59, 0.25}
\definecolor{YellowOrange}{cmyk}{0, 0.42, 1, 0}

\begin{document}

\begin{center}
{\bf\LARGE Conformal gravity with totally antisymmetric torsion} \\
\vskip 2 cm
{\bf \large Riccardo D'Auria$^{1}$, Lucrezia Ravera$^{1,2}$}
\vskip 8mm
 \end{center}
\noindent {\small $^{1}$ \it DISAT, Politecnico di Torino, Corso Duca degli Abruzzi 24, 10129 Torino, Italy. \\
    $^{2}$  \it INFN, Sezione di Torino, Via P. Giuria 1, 10125 Torino, Italy.
}

\vskip 2 cm
\begin{center}
{\small {\bf Abstract}}
\end{center}

We present a gauge theory of the conformal group in four spacetime dimensions with a non-vanishing torsion. In particular, we allow for a completely antisymmetric torsion, equivalent by Hodge duality to an axial vector whose presence does not spoil the conformal invariance of the theory, in contrast with claims of antecedent literature. The requirement of conformal invariance implies a differential condition (in particular, a Killing equation) on the aforementioned axial vector which leads to a Maxwell-like equation in a four-dimensional curved background. We also give some preliminary results in the context of $\mathcal{N}=1$ four-dimensional conformal supergravity in the geometric approach, showing that if we only allow for the constraint of vanishing supertorsion all the other constraints imposed in the spacetime approach are a consequence of the closure of the Bianchi identities in superspace. This paves the way towards a future complete investigation of the conformal supergravity using the Bianchi identities in the presence a non-vanishing (super)torsion.

\vfill
\noindent {\small{\it
    E-mail:  \\
{\tt riccardo.dauria@polito.it}; \\
{\tt lucrezia.ravera@polito.it}}}
   \eject


\section{Introduction}

In \cite{Kaku:1977pa} it was shown that the locally scale invariant Weyl theory of gravity is the gauge theory of the conformal group, where conformal transformations (conformal boosts) are gauged by a non-propagating gauge field.
In that theory the authors adopted the formalism of \cite{MacDowell:1977jt,Townsend:1977fz} to construct a quadratic Lagrangian with the curvatures associated with the conformal group in four spacetime dimensions. They claimed that in order to produce a conformally invariant theory in this setup it is necessary to set the torsion to zero.\footnote{Actually, in  \cite{Kaku:1977pa} the explicit form of the spin connection contains the dilaton, which gives a torsion trace contribution. However, the latter can be consistently set to zero in the theory, as already observed in \cite{Kaku:1977pa} and as we will also discuss in the present work. Therefore, let us refer to the theory of \cite{Kaku:1977pa} as a torsion-free one.}

In this work, in contrast with this claim, we show that it is actually possible to construct a gauge theory of the conformal group in four spacetime dimensions with a non-vanishing torsion component where proper conformal transformations are gauged by a non-propagating gauge field (the Schouten 1-form field).
In particular, we allow for a totally antisymmetric torsion, equivalent by Hodge duality to an axial vector, and still get a conformal gauge theory whose Lagrangian is quadratic in the curvatures of the conformal algebra (as the one of \cite{Kaku:1977pa}, that is the same construction of \cite{MacDowell:1977jt,Townsend:1977fz}).
We explicitly show how to reproduce the Weyl Lagrangian in this framework in the presence of a non-vanishing completely antisymmetric torsion and study the field equations of the theory. Let us also mention that for quadratic theories, in general, working in the first order or in the second order formalism for the spin connection one obtains different results. We will adopt the second order formalism, which will allow us to end up with a fourth order propagation equation for the graviton, \emph{the Lorentz connection being now torsionful}.

In this setup, invariance under conformal boosts (also known as proper, or special, conformal transformations) implies a \emph{Killing vector equation}, namely a differential condition on the axial vector torsion which, upon further differentiation, leads to a Maxwell-like equation in a four-dimensional curved background.
In the limit in which the torsion is set to zero we recover the conformal theory of \cite{Kaku:1977pa}.

The first part of the work will be devoted to study the purely bosonic gravitational theory. Subsequently, in view of a complete future investigation of the supersymmetric extension of this theory \cite{RL}, we give some preliminary results we have obtained regarding conformal supergravity. We will adopt the geometric approach to supergravity  (also called \emph{supergroup manifold} approach or \emph{rheonomic approach}).\footnote{For details on this formalism, see the original formulation in \cite{Castellani:1991et} and the pedagogical review \cite{DAuria:2020guc}.}

As pioneering works on the structure of conformal supergravity at the linearized level we refer the reader to \cite{Ferrara:1977mv,Ferrara:1978rk}.
The full conformal $\mathcal{N}=1$ supergravity theory in $D=3+1$ spacetime dimensions has been presented in \cite{Kaku:1978nz,Townsend:1979ki} (see also the review work \cite{Fradkin:1985am}) and subsequently rephrased in \cite{Castellani:1981um} in the geometric approach to supergravity. Interesting recent developments on $\mathcal N$-extended conformal supergravity and its spectrum in four dimensions have been recently obtained in Ref. \cite{Ferrara:2020zef}.
In all these papers, together with the vanishing of (super)torsion, a set of additional constraints were also imposed. In particular in Ref. \cite{Castellani:1981um} the constraints were implemented by the use of Lagrange multipliers. The constraint of vanishing supertorsion was justified arguing that only in this case the Lagrangian would have been invariant under special conformal transformations. Since we will prove in the sequel of this work that, at least at the purely bosonic level, one can still recover invariance under (special) conformal transformations allowing for a non-vanishing axial vector torsion, we argue that something similar should presumably happen in the superconformal case. 
In view of future investigations in this direction, here we start a preliminary analysis at the level of Bianchi identities in the geometric approach, showing that, besides the vanishing supertorsion, all the aforementioned constraints can be directly obtained from the study of the Bianchi identities,\footnote{See also Ref. \cite{Cribiori:2018xdy}, where the authors used the  geometric approach as in the present case, albeit in a different context, and the same conclusions can be reached after appropriate truncations.} just imposing the vanishing of the supertorsion (that is the supersymmetric extension of the constraint imposed in \cite{Kaku:1977pa}, where torsion was indeed assumed to vanish).

The paper is organized as follows: In Section \ref{reviewconf} we give a brief review of the conformal setup in four spacetime dimensions. In Section \ref{theory} we develop the gauge theory of the conformal group with a non-vanishing completely antisymmetric torsion. 
Subsequently, in Section \ref{prelres} we give some preliminary results regarding the extension to $\mathcal{N}=1$, $D=4$ conformal supergravity with vanishing supertorsion in the geometric approach. We conclude our work with some remarks and a discussion on future developments. In Appendix \ref{gammamatr} some useful formulas on gamma matrices in four dimensions are collected.

\section{Review of the gauging of the conformal group}\label{reviewconf}

The conformal group \cite{Barut:1970} $\rm{O}(4,2)$ is locally isomorphic to $\rm{SU}(2,2)$. The corresponding algebra is generated by the set of generators $\mathbf{T}_A=\lbrace \mathbf{J}_{ab}, \mathbf{P}_a, \mathbf{K}_a, \mathbf{D} \rbrace $, where we have decomposed the adjoint index $A$ of the conformal algebra with respect to the Lorentz indices $a,b,\ldots=0,1,2,3$. $\mathbf{J}_{ab}$ are the Lorentz rotations, $\mathbf{P}_a$ the spacetime translations, $\mathbf{K}_a$ the conformal boosts, and $\mathbf{D}$ the dilatation (scale transformation). In our conventions the metric $\eta_{ab}$ has signature $(+,-,-,-)$.\footnote{Regarding our conventions, throughout the paper we will use rigid Latin indices $a,b,\ldots=0,1,2,3$ instead of the world Greek indices $\mu, \nu, \ldots = 0,1,2,3$,  and will expand the $p$-forms in terms of the vierbein basis rather that in terms of differentials. For example, a generic 2-form will be expanded as ${F^A}={F^A}_{bc}V^bV^c={F^A}_{\mu \nu} d x^\mu \wedge dx^n$, where $V^a={V^a}_\mu dx^\mu$. This choice is convenient for the extension of the theory to superspace using the geometric formalism where the $p$-forms are expanded in terms of the full supervierbein basis $(V^a,\psi^\alpha)$, $\psi^\alpha$ being the gravitino 1-form. We will come back to a preliminary study of conformal supergravity in Section \ref{prelres}.}

Let us introduce the gauge 1-form fields $\omega^{ab}$ (spin connection\footnote{We call $\omega^{ab}$ the spin connection antisymmetric in $a,b$, $\omega^{ab}=-\omega^{ba}$, which may (and in fact will) involve torsion.}), $V^a$ (vierbein), $K^a$ (special conformal 1-form field), $\mathfrak{D}$ (dilaton gauge field), respectively dual to the vector fields generators of the conformal algebra, namely
\begin{equation}\label{dualgen}
\omega^{ab}(\mathbf{J}_{cd})=2\delta^{ab}_{cd}\,, \quad V^a(\mathbf{P}_b)=\delta^a_b \,, \quad K^a(\mathbf{K}_b)=\delta^a_b  \,, \quad \mathfrak{D}(\mathbf{D})= 1 \,.
\end{equation}
We can then write the corresponding curvatures,
\begin{equation}\label{curvatures}
\begin{split}
R^{ab} & \equiv \mathcal{R}^{ab} - 4 V^{[a} \wedge K^{b]} \,, \\
T^a & \equiv \mathcal{D} V^a + \mathfrak{D} \wedge V^a \,, \\
\mathcal{T}^a & \equiv \mathcal{D} K^a - \mathfrak{D} \wedge K^a \,, \\
G & \equiv d \mathfrak{D} + 2 V^a \wedge K_a \,,
\end{split}
\end{equation}
where $\mathcal{D}=d-\omega$ is the Lorentz covariant derivative and
\begin{equation}\label{lor}
\mathcal{R}^{ab}= d\omega^{ab}- {\omega^a}_c \wedge \omega^{cb}
\end{equation}
is the Riemann curvature.\footnote{We will generally omit writing of the wedge product between differential forms in order to lighten the notation.} 
Setting the curvatures \eqref{curvatures} to zero, the vanishing right-hand sides define the Maurer-Cartan equations, describing the ``vacuum'' (ground state), dual to the commutator algebra of the vector fields generators $\lbrace \mathbf{J}_{ab}, \mathbf{P}_a, \mathbf{K}_a, \mathbf{D} \rbrace $ (as it is well known, the $d^2$-closure of the Maurer-Cartan equations coincides with the Jacobi identities of the algebra).
For the sake of convenience, let us also define
\begin{equation}\label{weylderivative}
\begin{split}
\hat{\mathcal{D}} V^a & \equiv \mathcal{D} V^a + \mathfrak{D} \wedge V^a = dV^a - {\omega^a}_b V^b + \mathfrak{D} \wedge V^a  \,, \\
\hat{\mathcal{D}} K^a & \equiv \mathcal{D} K^a - \mathfrak{D} \wedge K^a = dK^a - {\omega^a}_b K^b - \mathfrak{D} \wedge K^a \,, 
\end{split}
\end{equation}
where $\hat{\mathcal{D}}$ denotes the \emph{Lorentz and scale} covariant differential.
The length-scale weight of the 1-forms and of their corresponding curvatures are 

\begin{equation}\label{formscale}
[\omega^{ab}]=[\mathfrak{D}]=0 \,, \quad [V^a]=1 \,, \quad [K^a]=-1 \,.
\end{equation}
Exploiting \eqref{weylderivative}, the curvatures \eqref{curvatures} can be recast into the following simpler expressions:
\begin{equation}\label{curvaturessimpler}
\begin{split}
R^{ab} & \equiv \mathcal{R}^{ab} - 4 V^{[a} K^{b]} \,, \\
T^a & \equiv \hat{\mathcal{D}} V^a \,, \\
\mathcal{T}^a & \equiv \hat{\mathcal{D}} K^a \,, \\
G & \equiv d \mathfrak{D} + 2 V^a K_a \,.
\end{split}
\end{equation}
and the Bianchi identities obeyed by the curvatures \eqref{curvaturessimpler} are

\begin{equation}\label{bianchiids}
\begin{split}
& \mathcal{D} R^{ab} + 4 (T^{[a} K^{b]} - V^{[a} \mathcal{T}^{b]}) =0 \,, \\
& \hat{\mathcal{D}} T^a + {R}^{ab}V_b - G V^a =0 \,, \\
& \hat{\mathcal{D}} \mathcal{T}^a + {R}^{ab} K_b + G K^a =0 \,, \\
& dG - 2 T^a K_a + 2 V^a \mathcal{T}_a =0 \,,
\end{split}
\end{equation}
where
\begin{equation}
\begin{split}
\hat{\mathcal{D}} T^a & \equiv \mathcal{D} T^a + \mathfrak{D} \wedge T^a \,, \\
\hat{\mathcal{D}} \mathcal{T}^a & \equiv \mathcal{D} \mathcal{T}^a - \mathfrak{D} \wedge \mathcal{T}^a \,.
\end{split}
\end{equation}
The conformal gauge transformations, associated with the conformal algebra, read
\begin{equation}\label{gaugetr}
\begin{split}
\delta \omega^{ab} & = \mathcal{D} \varepsilon^{ab} + 4 \varepsilon^{[a} K^{b]}-  4 V^{[a} \varepsilon^{b]}_K \,, \\
\delta V^a & = \hat{\mathcal{D}} \varepsilon^a + \varepsilon^{ab} V_b - \varepsilon_{\mathfrak{D}} V^a \,, \\
\delta K^a & = \hat{\mathcal{D}} \varepsilon^a_K + \varepsilon^{ab} K_b - \varepsilon_{\mathfrak{D}} K^a \,, \\
\delta \mathfrak{D} & = d \varepsilon_{\mathfrak{D}} - 2 \varepsilon^a K_a + 2 V_a \varepsilon^a_K \,,
\end{split}
\end{equation}
where $\varepsilon^{ab}$, $\varepsilon^a$, $\varepsilon^a_K$, and $\varepsilon_{\mathfrak{D}}$ are the Lorentz, translations, conformal boosts, and dilatation parameters, respectively.
Restricting ourselves to conformal boosts and dilatations we have
\begin{equation}\label{gaugetrrestricted}
\begin{split}
\delta \omega^{ab} & = - 4 V^{[a} \varepsilon^{b]}_K \,, \\
\delta V^a & = - \varepsilon_{\mathfrak{D}} V^a \,, \\
\delta K^a & = \hat{\mathcal{D}} \varepsilon^a_K - \varepsilon_{\mathfrak{D}} K^a \,, \\
\delta \mathfrak{D} & = d \varepsilon_{\mathfrak{D}} + 2 V_a \varepsilon^a_K \,.
\end{split}
\end{equation}

Let us mention that the theory whose Lagrangian we are going to consider (see eq. \eqref{lagr} in the following) is invariant under diffeomorphisms by construction, since it is written in terms of differential forms, but it is not invariant under spacetime translations. This is what commonly happens in gravitational theories. Thus, it is not a true ``gauge'' theory of the conformal group. However we shall adopt the terminology of ``gauge theory of the conformal group'' since it is widely used in the literature, keeping in mind that, in fact, we just have diffeomorphisms invariance rather than invariance under spacetime translations.\footnote{We recall that if we let the index $A$ denote the coadjoint representation, an infinitesimal diffeomorphism of anholonomic parameter $\varepsilon^A=\varepsilon^\rho \mu^A_\rho$ on any gauge field of the algebra $\mu^A$ can be written as $\delta_\varepsilon \mu^A = \mathrm{D} \varepsilon^A + \imath_\varepsilon R^A$, where $\mathrm{D}$ is the covariant derivative in the coadjoint representation. Therefore, the diffeomorphisms of the gauge fields differ from the gauge translations by a term proportional to the contraction of the curvature along an infinitesimal translation $\varepsilon^a \mathbf P_a$, where $\varepsilon^a$ is the infinitesimal parameter. In the supersymmetric case the contraction of the supercurvatures along a supersymmetric generator is also in general different from zero. Therefore, in the superconformal case the superspace translations correspond to supersymmetry transformations.}

Finally, let us also recall that the curvatures \eqref{curvaturessimpler} can be expanded along the vierbeins, which are dual to the spacetime translation generators. 
This amounts to the requirement of having conformal symmetry of a theory defined on spacetime and it is therefore a natural physical request to have a conformal gravity theory within our approach. 
Indeed, as Lorentz and scale symmetries are an exact invariance of the Lagrangian (which we will introduce in the following), the coset $$\frac{\mathrm{SU}(2,2)}{ \mathrm{SO}(1,3)\otimes \mathrm{O}(1,1)}$$ only depends on the vierbein and the $K^a$ 1-form. On the other hand, since we shall see that the gauge field $K^a$ can be expressed in terms of contractions of the Riemann tensor (more precisely, the Schouten tensor, as we will discuss in the sequel), the cotangent space is spanned in terms of the vierbein only.

Therefore the aforesaid expansion of the curvatures along the vierbein basis reads 
\begin{equation}\label{vierbasis}
\begin{split}
R^{ab} & = {R^{ab}}_{cd} V^c V^d \,, \\
T^a & = {T^a}_{bc} V^b V^c \,, \\
\mathcal{T}^a & = {T^a}_{bc} V^b V^c \,, \\
G & = G_{ab} V^a V^b \,.
\end{split}
\end{equation}
We will now proceed with the development of a gauge theory of the conformal group with a non-vanishing torsion.

\section{Gauge theory of the conformal group in the presence of a non-vanishing torsion}\label{theory}

We consider the same action introduced in \cite{Kaku:1977pa}, which is the only parity conserving quadratic action that can be constructed from the curvatures \eqref{curvaturessimpler} without dimensional constants,
\begin{equation}
\mathcal{A} = \int_{\mathcal M_4} \mathcal{L} \,,
\end{equation}
where
\begin{equation}\label{lagr}
\mathcal{L} = R^{ab} \wedge R^{cd} \epsilon_{abcd} 
\end{equation}
is the Lagrangian 4-form and $\mathcal M_4$ is the four-dimensional spacetime.

Let us first recall the well known fact that the variation of the action with respect to the special conformal 1-form $K^b$ gives an algebraic equation for the special conformal gauge field $K^a$. Indeed, varying the Lagrangian \eqref{lagr} with respect to $K^b$ we obtain the field equations
\begin{equation}
-8 V^a R^{cd} \epsilon_{abcd} =0 \,,
\end{equation}
which imply, using the expansion along the vierbein basis \eqref{vierbasis},
\begin{equation}\label{eomK}
\begin{split}
& R = 0 \,, \\
& \check{R}_{ab} =0 \,,
\end{split}
\end{equation}
where $R={R^{ab}}_{ab}$ and $\check{R}_{ab} ={R^c}_{acb}$. Taking \eqref{eomK} together with the definition of $R^{ab}$ in \eqref{curvaturessimpler} and writing $K_a=K_{ab}V^b$, one gets
\begin{equation}\label{KabSch}
K_{ab} = S_{ab} \,,
\end{equation}
being $S_{ab}$ the Schouten 0-form tensor defined in four-dimensional spacetime as
\begin{equation}\label{Sch}
S_{ab} \equiv \frac{1}{2} \left( \check{\mathcal{R}}_{ab} - \frac{1}{6} \eta_{ab}\mathcal{R} \right) \,,
\end{equation}
where $\mathcal{R}={\mathcal{R}^{ab}}_{ab}$ and $\check{\mathcal{R}}_{ab}$ are the scalar curvature and the Ricci tensor of $\omega^{ab}$, respectively. Notice that, in the presence of a non-vanishing torsion, $\omega^{ab}$ also includes a contorsion component implying that the Schouten tensor $S_{ab}$ has a non-vanishing antisymmetric part, $S_{[ab]}=\check{\mathcal{R}}_{[ab]}$, given entirely in terms of torsion.

Thus, we have obtained an algebraic equation for the non-propagating gauge field $K_{ab}$, namely eq. \eqref{KabSch}, which tells us that $K_{ab}$ corresponds to the Schouten tensor. This is a well known fact (see for instance \cite{Kaku:1977pa}). However, some comment is in order on this point, and we make it in the following. In particular, we will show that the fact that $K_{ab}$ corresponds to the Schouten can be actually deduced directly from a vacuum analysis and we will also give the irreducible decomposition of $\mathcal{R}_{abcd}$ which will be useful in the sequel.

\subsection{Curvature irreducible decomposition and Schouten tensor} 

Let us briefly discuss, before proceeding with our main results, that already at the vacuum level, as in fact expected, one can show that the field $K^a$ cannot be anything other than the Schouten 1-form,  $S^a=S^{ab}V_b$. This can be shown by taking into account the irreducible decomposition of the Riemann tensor ${\mathcal{R}^{ab}}_{cd}$ (here allowing also for the presence of torsion, see, for instance, \cite{Hehl:1994ue} for details).
Regarding the number of components, in four spacetime dimensions we have $\text{dim}\left({\mathcal{R}^{ab}}_{cd}\right)= 6 \times 6 = 20 \oplus 15 \oplus 1$, corresponding in terms of the {\rm {SL}}(4) representations to the dimensions of the following Young diagrams:

\begin{equation}\label{irrepsYoung}
\begin{ytableau}
$$ \cr $$
\end{ytableau}
\quad \otimes \quad
\begin{ytableau}
$$ \cr $$
\end{ytableau}
\quad = \quad
\begin{ytableau}
$$ & $$ \cr $$ & $$
\end{ytableau}
\quad \oplus \quad
\begin{ytableau}
$$ & $$ \cr $$ \cr $$ 
\end{ytableau}
\quad \oplus \quad
\begin{ytableau}
$$ \cr $$ \cr $$ \cr $$
\end{ytableau}
\end{equation}

Decomposing the $\mathrm{SL}(4)$ representations with respect to $\mathrm{SO}(1,3)$ in terms of their traceless plus trace parts  we find six irreducible pieces (irrepses):

\begin{equation}\label{irrepsYoung1}
\begin{ytableau}
$$ & $$ \cr $$ & $$
\end{ytableau}
\quad \longrightarrow \quad
\mathring{\begin{ytableau}
$$ & $$ \cr $$ & $$
\end{ytableau}
}
\quad \oplus \quad
 \mathring{\begin{ytableau}
$$ & $$
\end{ytableau}}
\quad \oplus \quad
\bullet
\end{equation}

\begin{equation}\label{irrepsYoung2}
\begin{ytableau}
$$ & $$\cr$$\cr$$
\end{ytableau}
\quad \longrightarrow \quad
\mathring{
\begin{ytableau}
$$ & $$ \cr $$\cr $$
\end{ytableau}
}
\quad \oplus \quad
 \mathring{
 \begin{ytableau}
$$ \cr $$
\end{ytableau}
}
\end{equation}

\begin{equation}\label{irrepsYoung3}
\begin{ytableau}
$$\cr $$\cr$$\cr$$\cr
\end{ytableau}
\quad \longrightarrow \quad
\begin{ytableau}
$$\cr $$\cr$$\cr$$\cr
\end{ytableau}
\end{equation}
where the small ring on top of the diagrams on the right-hand sides means that the corresponding representation is traceless, while the bullet denotes the scalar representation.

The three irrepses on the right-hand side of eq. \eqref{irrepsYoung1} correspond to the 10-dimensional Weyl tensor $W_{abcd}$, the  9-dimensional traceless symmetric Ricci tensor $\mathring{\check{\mathcal{R}}}_{(ab)}$, and the scalar curvature $\mathcal{R}$, respectively, and, using the nomenclature of Ref. \cite{Hehl:1994ue}, are called  WEYL + RICSIMF + SCALAR. The two irrepses on the right-hand side of eq. \eqref{irrepsYoung2} correspond to the 9-dimensional tensor which has the same number of degrees of freedom of a symmetric traceless tensor,\footnote{The corresponding representation is commonly referred to as \emph{associated}.} plus the 6-dimensional traceless antisymmetric Ricci tensor $\check{\mathcal{R}}_{[ab]}$, shortly referred together to as PAIRCOM and RICANTI, respectively. Finally, on the right-hand side of eq. \eqref{irrepsYoung3} we have a pseudo-scalar Hodge-dual to $\mathcal{R}_{[abcd]}$ denoted as PSSCALAR.

Writing
\begin{equation}
\begin{split}
\mathcal{R}_{abcd} \quad = \quad \underbrace{\frac{1}{2} \left( \mathcal{R}_{abcd} + \mathcal{R}_{cdab} \right)}_{\text{WEYL + RICSIMF + SCALAR}} \quad \oplus \quad \underbrace{\frac{1}{2} \left[ \left( \mathcal{R}_{abcd} - \mathcal{R}_{cdab} \right) - \mathcal{R}_{[abcd]} \right]}_{\text{PAIRCOM + RICANTI}} \quad \oplus \quad \underbrace{\mathcal{R}_{[abcd]}}_{\text{PSSCALAR}} \,,
\end{split}
\end{equation}
the three ``underbraced'' expressions correspond to the left-hand sides of the three 
eqs. \eqref{irrepsYoung1}, \eqref{irrepsYoung2}, and \eqref{irrepsYoung3}, respectively.
We enumerate the various representatios writing 
\begin{equation}\label{irrepsR}
\begin{split}
{\mathcal{R}^{ab}}_{cd} & = \sum_{i=1}^6 {\mathcal{R}^{ab}}_{(i)|cd} \\
& = \underbrace{{\mathcal{R}^{ab}}_{(1)|cd}}_{\text{WEYL [10]}} + \underbrace{{\mathcal{R}^{ab}}_{(2)|cd}}_{\text{PAIRCOM [9]}} + \underbrace{{\mathcal{R}^{ab}}_{(3)|cd}}_{\text{PSSCALAR [1]}}  + \underbrace{{\mathcal{R}^{ab}}_{(4)|cd}}_{\text{RICSIMF [9]}} + \underbrace{{\mathcal{R}^{ab}}_{(5)|cd}}_{\text{RICANTI [6]}} + \underbrace{{\mathcal{R}^{ab}}_{(6)|cd}}_{\text{SCALAR [1]}} \,,
\end{split}
\end{equation}
where the numbers in the square brackets denote the components of each irrep (dimension of the irrep).

Let us observe that ${\mathcal{R}^{ab}}_{(2)|cd}$, ${\mathcal{R}^{ab}}_{(3)|cd}$, and ${\mathcal{R}^{ab}}_{(5)|cd}$ are non-vanishing only in the presence of torsion (so that they are given in terms of torsion and its derivatives).

Now, we can exploit \eqref{irrepsR} in the vacuum of our theory,  given by the vanishing right-hand side of \eqref{curvaturessimpler}, In particular we have

\begin{equation}
\mathcal{R}^{ab} - 4 V^{[a} K^{b]} = 0 \,,
\end{equation}
that is
\begin{equation}\label{vacuumeqKabSab}
{\mathcal{R}^{ab}}_{cd} - 4 {\delta^{[a}}_{[c} {K^{b]}}_{d]} = 0 \,.
\end{equation}
Using \eqref{irrepsR} and observing that the second term in \eqref{vacuumeqKabSab} can be written only in terms of the irreducible pieces RICSYMF, SCALAR, and RICANTI, one can prove that $K_{ab}$ must coincide with the Schouten tensor $S_{ab}$ (that is, $K^a={S^a}_b V^b$), the components of the latter being defined in \eqref{Sch}.\footnote{Indeed, tracing the $b,d$ indices of both terms on the left-hand side of eq. \eqref{vacuumeqKabSab} one easily recover $K_{ab}=S_{ab}$.}
Therefore, the Maurer-Cartan equations obtained from \eqref{curvaturessimpler} take the following form:
\begin{equation}\label{vacuumnew}
\begin{split}
\mathcal{R}^{ab}_{(2)}&\equiv{\mathcal{R}^{ab}}_{(2)|cd}V^cV^d=0 \,,  \\
\mathcal{R}^{ab}_{(3)}&\equiv{\mathcal{R}^{ab}}_{(3)|cd}V^cV^d=0 \,, \\     
W^{ab} &\equiv \mathcal{R}^{ab}_{(1)} = \mathcal{R}^{ab} - 4 V^{[a} S^{b]} = 0 \,, \\
T^a &\equiv\hat{\mathcal{D}} V^a=0 \,, \\ 
\mathcal{T}^a & \equiv \hat{\mathcal{D}} S^a =0 \,,\\ 
G & \equiv d \mathfrak{D} + 2 V^a S_a = 0 \,,
\end{split}
\end{equation}
where $\mathcal{R}^{ab}_{(2)}$ and $\mathcal{R}^{ab}_{(3)}$ are the PAIRCOM and PSCALAR 2-forms, respectively, and $W^{ab}={W^{ab}}_{cd}V^cV^d$, being ${W^{ab}}_{cd}$ the Weyl tensor.

One could then go out of the vacuum switching on the curvatures associated with the Maurer-Cartan equations \eqref{vacuumnew} and write a quadratic Lagrangian in terms of these field-strengths.
We note that in the case of vanishing torsion ($T^a=0 \Rightarrow \mathcal{R}^{ab}_{(2)}=\mathcal{R}^{ab}_{(3)}=0 $) such Lagrangian reads
\begin{equation}\label{WW}
\mathcal{L}_W = W^{ab}\wedge W^{cd} \epsilon_{abcd} \,,
\end{equation}
where $W^{ab}=W^{ab}(\omega)$ with $\omega=\omega(\mathfrak{D},V)$ (note that we still have ${\mathcal{R}^{ab}}_{(5)|cd}\neq 0$, due to the presence of $\mathfrak{D}$). The Lagrangian \eqref{WW} coincides with the Lagrangian given in Ref. \cite{Kaku:1977pa}. In that paper the authors set  the torsion $T^a$ equal to zero right from the beginning, and  plugging back into \eqref{lagr} the on-shell expression \eqref{KabSch} for $K_{ab}$, they recover the same Weyl Lagrangian \eqref{WW}. As it was to be expected, the theory \eqref{WW} that we have constructed by gauging directly the Maurer-Cartan equations \eqref{vacuumnew} coincides, when $T^a$ is zero, with the theory of Ref. \cite{Kaku:1977pa}. In fact, eq. \eqref{KabSch} shall be interpreted directly as a consequence of the structure of the vacuum  of the theory quadratic in the Weyl tensor, which is indeed the conformal theory we are going to focus on. 

In the sequel we will show that, remarkably, the Lagrangian we will develop describing a gauge theory of the conformal group  with a \emph{non-vanishing torsion} is formally identical to the Lagrangian \eqref{WW} provided  the curvatures are constructed from a torsionful  connection.
In view of this, let us proceed by first showing that it is still possible to  get conformal invariance of the theory in the presence of a non-vanishing $T^a$. In other words, we are going to prove that the constraint of vanishing torsion introduced in \cite{Kaku:1977pa} to get a conformally invariant theory can be actually relaxed. 


\subsection{Conformal invariance of the theory}

The aim of this section is to see whether a non-vanishing $T^a$ is allowed in a ``gauge'' theory of the conformal group. In particular, we will allow for a totally antisymmetric torsion, equivalent by Hodge duality to an axial vector.
We will show that the requirement of conformal invariance of the Lagrangian \eqref{lagr} constructed with the curvatures \eqref{curvaturessimpler} can be still fulfilled provided we require the vanishing of $R^{ab}_{(2)}$, $R^{ab}_{(3)}$, and $R^{ab}_{(5)}$ (notations for the irrepses of $R^{ab}$ are the same as for $\mathcal{R}^{ab}$). This will imply a differential condition on the completely antisymmetric part of the torsion, namely for the aforesaid Hodge-dual axial vector. Upon use of the on-shell conditions \eqref{eomK} implying \eqref{KabSch}, the theory will \emph{formally} reproduce the same Lagrangian \eqref{WW}, albeit with a torsionful spin connection.\footnote{Note that the same Lagrangian can be also obtained by directly gauging the Maurer-Cartan equations \eqref{vacuumnew}, that is switching on the corresponding curvatures going out of the vacuum, in the presence of a completely antisymmetric torsion, and with the aforementioned constraints.}
 
Let us show the above explicitly.  
The Lagrangian \eqref{lagr} is clearly scale invariant, as $R^{ab}$ has zero scale weight. Nevertheless, the invariance under conformal boosts can be achieved only in a non-trivial way. In particular, one can prove that some constraints on the curvatures \eqref{curvaturessimpler} have to be imposed in order for \eqref{lagr} to be invariant under proper conformal transformations on spacetime. Indeed, to recover invariance of \eqref{lagr} under conformal boosts, performing the variation $\delta K^d = \hat{\mathcal{D}} \varepsilon^d_K$ (see \eqref{gaugetr}), we must have
\begin{equation}
\left( R^{ab} \epsilon_{abcd} T^c \right) \varepsilon^d_K = 0 \,,
\end{equation}
that is, using \eqref{vierbasis},
\begin{equation}\label{eqconfinv}
\left( {R^{ab}}_{lm} {T^c}_{pq} \epsilon^{lmpq} \epsilon_{abcd} \Omega^{(4)} \right) \varepsilon^d_K = 0 \,,
\end{equation}
where the $\Omega^{(4)}$ is the four-dimensional volume element defined as $\Omega^{(4)}\equiv- \frac{1}{4!} \epsilon_{abcd}V^a V^b V^c V^d$.
In \cite{Kaku:1977pa} the authors claimed that \eqref{lagr} results to be invariant under proper conformal gauge transformations only if $T^a=0$ (the vanishing of $ {R^{ab}}_{lm}$ being not considered as it would trivialize the theory). Actually, this is not the case, as we will show in the sequel.

In order to explain our claim in detail we  need the irreducible decomposition of the torsion in four spacetime dimensions (see, for instance, \cite{Hehl:1994ue,Gronwald:1995em,Klemm:2018bil}).
In four dimensions the torsion tensor ${T^a}_{bc}$ has $24=20 \oplus 4$ components, and we may write its decomposition as
\begin{equation}\label{irrepsTYoung}
\begin{ytableau}
$$ $$
\end{ytableau}
\quad \otimes \quad
\begin{ytableau}
$$ \cr $$
\end{ytableau}
\quad = \quad
\mathring{\begin{ytableau}
$$ $$ & $$ \cr $$ 
\end{ytableau}}
\quad \oplus \quad
\begin{ytableau}
$$ $$
\end{ytableau}
\quad \oplus \quad
\begin{ytableau}
$$ \cr $$ \cr $$ $$
\end{ytableau} 
\end{equation}
whose dimensions are 16, 4, and 4, respectively, corresponding to the decomposition
\begin{equation}
{T^a}_{bc}=T_{\mathring{\begin{ytableau}
b$$ &a $$ \cr c$$ $$
\end{ytableau}}
}
+\frac23\delta^a_{[b} t_{c]}+T_{[abc]} \,.
\end{equation}
In the following we will denote the 16-dimensional representation as a tensor ${Z^a}_{bc}$, that is
\begin{equation}
T_{\mathring{\begin{ytableau}
b$$ &a $$ \cr c$$ $$
\end{ytableau}}
}={Z^a}_{bc} \,,
\end{equation}
and the antisymmetric representation $T_{[abc]}$ as the axial vector $\tilde t^d$, namely
\begin{equation}
T_{[abc]}=-\frac{1}{6}\epsilon_{abcd} \tilde t^d \,,
\end{equation}
while $t^a$ appearing in the torsion trace part is an ordinary vector. 
Inserting the above decomposition of the torsion into \eqref{eqconfinv}, the latter becomes
\begin{equation}\label{eqconfinvcomplete}
\left[ {R^{ab}}_{lm} \left(\frac{2}{3} \delta^c_p t_q - \frac{1}{6} \epsilon^{cpqr} \tilde{t}_r + {Z^c}_{pq} \right) \epsilon^{lmpq} \epsilon_{abcd} \right] \varepsilon^d_K = 0 \,.
\end{equation}
The necessary condition given by \eqref{eqconfinvcomplete} consists in a set of four
algebraic equations (recall that $\varepsilon^d_K$ is arbitrary) in the curvatures $R^{ab}$ and ${T^a}_{bc}$ for which we are now going to 
examine in detail some particular solutions. 

We first observe that a sufficient constraint on the torsion to have conformal boosts invariance is $\tilde{t}_a={Z^a}_{bc}=0$. Indeed, in this case \eqref{eqconfinvcomplete} yields
\begin{equation}
\left( - R t_a + 2 \check{R}_{ba} t^b \right) \varepsilon^a_K = 0 \quad \longrightarrow \quad R t_a - 2 \check{R}_{ba} t^b = 0 \,,
\end{equation} 
which, for a non-vanishing torsion trace $t_a$, has as a particular solution
\begin{equation}
R=0 \,, \quad \check{R}_{ab}=0 \,.
\end{equation}
The latter constraints coincide with the equations that one obtains when varying the Lagrangian \eqref{lagr} with respect to $K^a$ (that is, when going on-shell for $K^a$), namely with \eqref{eomK}. 
However, let us recall here that the torsion trace $t_a$, even if perfectly allowed, as we have just seen that it does not spoil the conformal invariance of the theory, can be actually set to zero in a consistent way (see e.g. \cite{Howe:2020hxi} for details). Indeed, one can easily verify that the torsion is invariant under a shift of the dilaton $\mathfrak{D}=\mathfrak{D}_a V^a$ by a parameter $X_a$, namely $\mathfrak{D}_a \rightarrow \mathfrak{D}'_a = \mathfrak{D}_a + X_a$, provided that the spin connection $\omega_{ab|m}=\omega_{ab|\mu} V^\mu_m$ transforms as
${\omega^{ab}}_{|m} \rightarrow {\omega'^{ab}}_{|m} = {\omega^{ab}}_{|m} - 2 {\delta_m}^{[a} X^{b]}$, and with the choice $X_a=\frac{2}{3}t_a$ the torsion trace gets reabsorbed into the dilaton (we have $\mathfrak{D}'_a = \mathfrak{D}_a + \frac{2}{3}t_a$, and $\mathfrak{D}'_a$ will be again renamed as $\mathfrak{D}_a$ in the following).
Thus, from now on we set
\begin{equation}\label{tazero}
t_a=0 \,.
\end{equation}
It follows that the necessary condition for conformal invariance \eqref{eqconfinvcomplete} becomes
\begin{equation}\label{necessary}
\left[ {R^{ab}}_{lm} \left({Z^c}_{pq} - \frac{1}{6} \epsilon^{cpqr} \tilde{t}_r \right) \epsilon^{lmpq} \epsilon_{abcd} \right] \varepsilon^d_K = 0 \,.
\end{equation}
A simple solution of eq. \eqref{necessary} can be found assuming
\begin{equation}\label{Zzero}
{Z^c}_{pq}=0 \,.
\end{equation}
Using \eqref{Zzero}, eq. \eqref{necessary} becomes
\begin{equation}\label{necessaryrestr}
 2 \epsilon^{bdlm} {R^a}_{dlm} \tilde{t}_a - \epsilon^{cdlm} R_{cdlm} \tilde{t}^b = 0 \,.
\end{equation}
We will now show that \eqref{Zzero} and the ensuing condition in \eqref{necessaryrestr} lead to intriguing physical consequences on the surviving field $\tilde t_a$. Indeed, for $\tilde{t}_a \neq 0$ a possible solution of \eqref{necessaryrestr} is\footnote{Here let us mention that, in fact, using \eqref{Zzero} into \eqref{necessary} the latter boils down to $${R^{ab}}_{lm} \tilde{t}_r \epsilon^{cpqr} \epsilon^{lmpq} \epsilon_{abcd} = 0 \,,$$ which can be simplified by contracting either first $\epsilon^{cpqr} \epsilon_{lmpq}$ and then the result with $\epsilon_{abcd} $, or first $\epsilon^{cpqr} \epsilon_{abcd}$ and then the result with $\epsilon^{lmpq}$. The respectively obtained equations may appear different at first sight, but exploiting the symmetry properties of the irrepses of $R_{abcd}$ one can verify that they are actually equivalent, both exhibiting, in particular, \eqref{goodsol} as a possible solution.} \begin{equation}\label{goodsol}
R_{a[bcd]}=0 \,, \quad R_{[abcd]} =0 \,.
\end{equation}
We shall focus on this particular solution. The latter implies 
\begin{equation}\label{ttneq0irr}
{R^{ab}}_{(2)|cd} + {R^{ab}}_{(5)|cd} = 0 \,, \quad {R^{ab}}_{(3)|cd} = 0 \,.
\end{equation}
Furthermore, recalling eq. \eqref{eomK}, namely 
\begin{align}
&\check{R}_{ab}= R=0 \quad \longleftrightarrow \quad {R^{ab}}_{(4)|cd} = {R^{ab}}_{(5)|cd} = {R^{ab}}_{(6)|cd}=0 \,,
\end{align}
and \eqref{KabSch}, plugging all of this into the definition of $R^{ab}$ in \eqref{curvaturessimpler}, one can easily realize that we are left with
\begin{equation}
\begin{split}
& {R^{ab}}_{(2)|cd} = 0 \quad \Rightarrow \quad {\mathcal{R}^{ab}}_{(2)|cd} = 0 \,, \\
& {R^{ab}}_{(3)|cd} = 0 \quad \Rightarrow \quad {\mathcal{R}^{ab}}_{(3)|cd} = 0 \,, \\
& {R^{ab}}_{(4)|cd} = 0 \,, \quad {R^{ab}}_{(5)|cd} = 0 \,, \quad {R^{ab}}_{(6)|cd} = 0 \,, 
\end{split}
\end{equation}
together with
\begin{equation}
\begin{split}
& {\mathcal{R}^{ab}}_{(4)|cd} + {\mathcal{R}^{ab}}_{(5)|cd} + {\mathcal{R}^{ab}}_{(6)|cd} \equiv 4 {\delta^{[a}}_{[c} {S^{b]}}_{d]} \,, \\
& {R^{ab}}_{(1)|cd} = {\mathcal{R}^{ab}}_{(1)|cd} \equiv {W^{ab}}_{cd} \,.
\end{split}
\end{equation}
Hence, since now we have
\begin{equation}
{R^{ab}}_{cd}={R^{ab}}_{(1)cd}={W^{ab}}_{cd} \equiv {\mathcal{R}^{ab}}_{cd} - \sum_{i=2}^6 {\mathcal{R}^{ab}}_{(i)|cd} = {\mathcal{R}^{ab}}_{cd} - 4 {\delta^{[a}}_{[c} {S^{b]}}_{d]} \,,
\end{equation}
we may write
\begin{equation}\label{RW}
\begin{split}
R^{ab} = W^{ab} & \equiv \mathcal{R}^{ab} - 4 V^{[a} S^{b]} \\
& = \mathcal{R}^{ab} - 2 \check{\mathcal{R}}^{[a|c}V_c V^{b]} + \frac{1}{3}\mathcal{R} V^a V^b \,,
\end{split}
\end{equation}
which is formally identical to the torsionless $W^{ab}$, but now \emph{$\omega^{ab}$ contains a torsion part}.

In conclusion, we have recovered invariance under conformal boosts of \eqref{lagr} solving the necessary condition for conformal invariance \eqref{necessary} under the assumption ${Z^a}_{bc}=0$, the only non vanishing part of the torsion $T^a$ being given by
\begin{equation}\label{torsnow}
T^a = {T^a}_{bc} V^b V^c = - \frac{1}{6} \epsilon^{abcd} \tilde{t}_d V^b V^c \,.
\end{equation} 
Inserting the actual form \eqref{torsnow} into
eq. \eqref{curvaturessimpler} we are led to
\begin{equation}\label{omegaexpr}
\omega_{ab|m} = \omega_{ab|\mu} V^\mu_m = \mathring{\omega}_{ab|m}- 2 \eta_{m[a} \mathfrak{D}_{b]} - \frac{1}{6} \epsilon_{abmc} \tilde{t}^c \,,
\end{equation}
where $\mathring{\omega}_{ab|m} = \mathring{\omega}_{ab|\mu} V_m^\mu$ and the last term in \eqref{omegaexpr} is the contribution due to the contorsion term.\footnote{Recall also that $\mathring{\omega}_{ab|\mu}  = \left( f_{\lambda|\mu\nu} + f_{\nu|\lambda \mu} - f_{\mu|\nu \lambda} \right) V_a^\lambda V_b^\nu$, with $f_{\lambda|\mu\nu} = V_\lambda^k \partial_{[\mu} V_{\nu]}^c \eta_{ck}$.}
Moreover, from the variation of the torsion definition in \eqref{curvaturessimpler}, we now have 
\begin{equation}\label{deltaom}
\begin{split}
\delta \omega_{ab|m} & = \left( \delta^l_m \delta^q_{[a} \delta^p_{b]} + \delta^q_m \delta^p_{[a} \delta^l_{b]} - \delta^p_m \delta^l_{[a} \delta^q_{b]} \right) \hat{\mathcal{D}}_q \left( \delta V^p \right)_l - 2 \eta_{m[a} \delta^c_{b]} \delta \mathfrak{D}_c \\
& - \frac{1}{6} \epsilon_{abmc} \delta \tilde{t}^c - \frac{1}{6} \epsilon_{abpc} \tilde{t}^c \delta^l_m ( \delta V^p )_l \,,
\end{split}
\end{equation}
and one can verify that the transformations of the fields in \eqref{deltaom} are such that the variation of $\omega_{ab|m}$ under dilatations and conformal boosts is the same whether one determines it from the gauge prescription \eqref{gaugetr} or directly from \eqref{deltaom}.\footnote{Indeed, under dilatations we have $\delta_{\varepsilon_{\mathfrak{D}}} {V^p}_l = - \varepsilon_{\mathfrak{D}} \delta^p_l$, $\delta_{\varepsilon_{\mathfrak{D}}} \mathfrak{D}_c = \partial_c \varepsilon_\mathfrak{D} = \hat{\mathcal{D}}_c \varepsilon_\mathfrak{D}$, and $\delta_{\varepsilon_{\mathfrak{D}}} \tilde{t}^c=\varepsilon_{\mathfrak{D}} \tilde{t}^c$, which plugged into \eqref{deltaom} lead to $\delta_{\varepsilon_{\mathfrak{D}}} \omega_{ab|m} (V,\mathfrak{D},\tilde{t})=0$ (in particular, the torsion contributions cancel each other out), reproducing the same result that can be obtained from the gauge prescription \eqref{gaugetr}. Analogously, under conformal boosts one has $\delta_{\varepsilon_K} {V^p}_l=0$, $\delta_{\varepsilon_K} \mathfrak{D}_c = 2 \varepsilon_{K|c}$, and $\delta_{\varepsilon_K} \tilde{t}^c=0$, implying $\delta_{\varepsilon_K} \omega_{ab|m} (V,\mathfrak{D},\tilde{t})= - 4 \eta_{m[a} \varepsilon_{K|b]} $, which is the same variation that one can obtain from \eqref{gaugetr}.} 
Thus, the Lagrangian \eqref{lagr} remains scale and proper conformal invariant if $\omega = \omega(V,\mathfrak{D},\tilde{t})$ as given in \eqref{omegaexpr}.

If we now substitute \eqref{RW} into \eqref{lagr}, we obtain a Lagrangian that is formally identical to \eqref{WW} but involving torsion, namely
\begin{equation}\label{weyllagr}
\begin{split}
\mathcal{L} & = W^{ab} W^{cd} \epsilon_{abcd} = -  W^{abcd} W_{abcd} \Omega^{(4)} \\
& = \mathcal{R}^{ab} \mathcal{R}^{cd} \epsilon_{abcd} - 8 \left( \check{\mathcal{R}}_{ab} \check{\mathcal{R}}^{ba} - \frac{1}{3} \mathcal{R}^2 \right) \Omega^{(4)} \,,
\end{split}
\end{equation}
the torsion being now hidden in the torsionful spin connection. 

Furthermore, at the level of the theory \eqref{lagr} the dilaton can be consistently eliminated. Indeed, also in the present case (as already pointed out in \cite{Kaku:1977pa} in the case of vanishing torsion) the kinetic term for $\mathfrak{D}$ does not contribute, since in \eqref{weyllagr} we have the combination $\check{\mathcal{R}}_{ab}\check{\mathcal{R}}^{ba}$ rather than $\check{\mathcal{R}}_{ab}\check{\mathcal{R}}^{ab}$ (whose presence would instead yield such kinetic term for the dilaton), which makes a difference because $\check{\mathcal{R}}_{ab}(\omega)$ is not symmetric. A further check of the non-propagating nature of the dilaton can be ascertained from its equation of motion. Indeed, even if we retain $\mathfrak{D}$ in the Lagrangian, one can verify that its equation of motion is actually the trivial identity. With these arguments, one may set\footnote{Notice that, in the present case, setting $\mathfrak{D}=0$ we still have a \emph{torsionful spin connection}, as the torsion axial vector $\tilde{t}^a$ does not vanish.} 
\begin{equation}\label{Dzero}
\mathfrak{D}=0
\end{equation}
from the start, but it is not immediately obvious how the Lagrangian remains invariant in this case since now the variation of $\omega_{ab|m}$ under dilatations and conformal boosts determined from \eqref{deltaom} is not the same as the one determined from the gauge prescription \eqref{gaugetr} anymore. This problem was solved in Ref. \cite{Kaku:1977pa} (for the torsion-free theory) showing that the additional terms present in the $\delta\omega^{ab}$ variation give a vanishing contributuion. The same mechanism holds true in our case. Indeed, when varying the Lagrangian we get an additional variation $\delta ' \mathcal{L}= \frac{\delta \mathcal{L}}{\delta \omega} \delta\omega'$, where $\delta\omega'$ is the difference between the gauge variation
of $\omega$ and the variation found from the explicit form of $\omega=\omega(V, \tilde{t})$. In particular, we get
\begin{equation}\label{deltapL}
\begin{split}
\phantom{\Rightarrow} \quad & \delta ' \mathcal{L} = - 16 \left(\frac{1}{6} \epsilon^{clnr} \tilde{t}_r {K^d}_p - \delta^c_l {\mathcal{T}^d}_{np} \right) \epsilon^{lnpq} \epsilon_{qbcd} \xi^b \\
\Rightarrow \quad & \delta ' \mathcal{L} = 16 \left( \frac{1}{3} \epsilon_{blnp} K^{ln} \tilde{t}^p + 4 {\mathcal{T}^l}_{bl} \right) \xi^b \,,
\end{split}
\end{equation}
where $\xi_b$ is either $\partial_b \varepsilon_{\mathfrak{D}}$ for dilations or $2 \varepsilon^c_K$ for conformal boosts. However, from the Bianchi identity of $R^{ab}$, which yields, in particular, ${\mathcal{T}^a}_{la}=-\frac{1}{12} \epsilon_{labn} K^{ab} \tilde{t}^n$, one can see that $\delta ' \mathcal{L}$ in \eqref{deltapL} vanishes identically. Hence, one can set the dilaton to zero without spoiling the conformal invariance of the theory, and \eqref{weyllagr}, that is Weyl gravity with a non-vanishing axial vector torsion, results to be the gauge theory of the conformal group with non-vanishing torsion. 

Having eliminated the dilaton, we are left with \eqref{RW} and \eqref{weyllagr} where now the dilaton contributions vanish.
In particular, now we have
\begin{equation}\label{TGnew}
G = 2 V^a S_a \quad \Rightarrow \quad G_{ab} = \check{\mathcal{R}}_{[ab]} \,.
\end{equation}   
Recalling that $\check{\mathcal{R}}_{[ab]}$ is a function of $\tilde{t}_a$ only and, in particular, through \eqref{omegaexpr} we have
\begin{equation}
\check{\mathcal{R}}_{[ab]}=- \frac{1}{12} \epsilon_{abcd} \mathcal{D}^c \tilde{t}^d \,,   
\end{equation}
eq. \eqref{TGnew} becomes
\begin{equation}\label{Gabfinal}
G_{ab} = \check{\mathcal{R}}_{[ab]} = - \frac{1}{12} \epsilon_{abcd} \mathcal{D}^c \tilde{t}^d \quad \longrightarrow \quad \check{\mathcal{R}}_{[ab]} = - \frac{1}{6} {}^\star \mathbb{T}_{ab} \,,
\end{equation}
where the star symbol denotes the Hodge duality operator and we have defined the field-strength
\begin{equation}\label{fieldstrengthtt}
\mathbb{T}_{ab} \equiv \mathcal{D}_{[a} \tilde{t}_{b]}  \,.
\end{equation}
We note that this result directly follows from the conformal invariance of the theory.

Furthermore, let us observe that the conformal invariance constraints in \eqref{goodsol} imply
\begin{align}
& \mathcal{R}_{a[bcd]}=-2 \eta_{a[d}S_{bc]} \quad \longrightarrow \quad \mathcal{R}_{a[bcd]}=-\eta_{a[d}\check{\mathcal{R}}_{bc]} \,, \label{confconstr1} \\
& \epsilon^{abcd}\mathcal{R}_{abcd}=0 \,. \label{confconstr2}
\end{align}
We now show that eqs. \eqref{confconstr1} and \eqref{confconstr2} imply a differential constraint on the axial vector part of the torsion. Indeed, from eq. \eqref{Gabfinal}, recalling that now we have $\omega=\omega(V,\tilde{t})$, that is the spin connection also involves a contorsion part, one can show that \eqref{confconstr2} reduces to
\begin{equation}\label{eqsfortorsfromconfinv1}
\mathcal{D}^a \tilde{t}_a = 0 \,.    
\end{equation}
Then, in \eqref{confconstr1} we express both sides of the equation in terms of the torsion by exploiting the fact that now the curvature and antisymmetric Ricci tensors are given entirely in terms of the totally antisymmetric torsion. Indeed, recall that the connection $\omega_{ab|c}$ involves, besides the usual Riemannian part, also a contorsion term $\mathcal{K}_{abc}$ related to the completely antisymmetric torsion as follows:
\begin{equation}
V^b {\mathcal{K}^a}_{b} = T^a \quad \Rightarrow \quad \mathcal{K}_{abc} = - \frac{1}{6} \epsilon_{abcd} \tilde{t}^d \,.    
\end{equation}
As a consequence, using \eqref{eqsfortorsfromconfinv1} into the explicit form of \eqref{confconstr1}, one is left with
\begin{equation}\label{eqsfortorsfromconfinv2}
\mathcal{D}_{(a} \tilde{t}_{b)} = 0 \,,
\end{equation}
that is a Killing equation for the axial vector $\tilde{t}_a$ of the present conformal theory as \eqref{eqsfortorsfromconfinv1} holds.
Hence, from \eqref{goodsol} we obtain eqs. \eqref{eqsfortorsfromconfinv1} and \eqref{eqsfortorsfromconfinv2} for the axial vector torsion $\tilde{t}_a$. These imply, together, as can be proven by further differentiation and using the fact that $\left[\mathcal{D}_a,\mathcal{D}_b\right] \tilde{t}^m= -2 {\mathcal{R}^m}_{nab}\tilde{t}^b -2 {T^n}_{ab} \mathcal{D}_n \tilde{t}^m = -2 {\mathcal{R}^m}_{nab}\tilde{t}^n  + \frac{1}{3} \epsilon_{abnd} \tilde{t}^d \mathcal{D}^n \tilde{t}^m$,
\begin{equation}\label{proptors}
\square \tilde{t}_b -2 \check{\mathcal{R}}_{ab} \tilde{t}^a - \frac{1}{3} \epsilon_{abcd} \tilde{t}^a \mathbb{T}^{cd} = 0 \quad \longrightarrow \quad \square \tilde{t}_b -2 \check{\mathcal{R}}_{ab} \tilde{t}^a + 4 \check{\mathcal{R}}_{[ab]} \tilde{t}^a  = 0 \,,
\end{equation}
where $\square$ denotes the covariant d'Alembertian with torsion, $\square \tilde{t}_b \equiv \mathcal{D}^a \mathcal{D}_a \tilde{t}_b$. 

Thus, as we can see from \eqref{proptors}, in our theory the axial vector torsion obeys \eqref{proptors}, and this follows directly from the requirement of invariance of the Lagrangian under conformal boosts. Eq. \eqref{proptors} can be regarded as a Maxwell-like equation in a curved four-dimensional background.\footnote{Observe that the field-strength $\mathbb{T}_{ab}$ of the axial vector actually determines the antisymmetric part of the Ricci tensor of $\omega$ as given by \eqref{Gabfinal}. In particular, the latter means that for the antisymmetric part of the Schouten in \eqref{KabSch} and \eqref{Sch} we have $S_{[ab]}=\check{\mathcal{R}}_{[ab]}=- \frac{1}{6} {}^\star \mathbb{T}_{ab}=- \frac{1}{12} \epsilon_{abcd} \mathcal{D}^c \tilde{t}^d$, where the (torsionful) covariant derivative $\mathcal{D}$ here, specifically, contributes only with the purely Levi-Civita part, since the torsion term in $\mathcal{D}$ cancels in this expression. Allowing also for a trace part of the torsion one would get, besides, bilinear torsion terms, but here we have already shown that the torsion trace can be consistently set to zero. Moreover, let us mention that the axial vector $\tilde{t}^a$ also contributes in a non-trivial way to the symmetric part of the Schouten tensor, that is it contributes to $S_{(ab)}$ with a term proportional to $\tilde{t}_a \tilde{t}_b$.}
Note, however, that this is not an equation of motion derived from the Lagrangian, but just the result of having required the conformal invariance in the presence of an axial vector torsion.

As a last comment let us observe that, taking into account all the results obtained till now, the Bianchi identities \eqref{bianchiids} can be rewritten as
\begin{equation}\label{Bianchifinal}
\begin{split}
\text{Bianchi for } R^{ab} \, \, (R^{ab} = W^{ab}) \, &: \quad \mathcal{D} W^{ab} + 4 \left( T^{[a} S^{b]} - V^{[a} C^{b]} \right) = 0 \quad \longleftrightarrow \quad \mathcal{D} \mathcal{R}^{ab} = 0 \,, \\
\text{Bianchi for } T^{a} \, &: \quad \mathcal{D} T^a + W^{ab} V_b - G V^a = 0 \,, \\
\text{Bianchi for } \mathcal{T}^{a} \, \, (\mathcal{T}^{a} = C^{a}) \, &: \quad \mathcal{D} C^a + \mathcal{R}^{ab} S_b = 0 \,, \\
\text{Bianchi for } G \, &: \quad d G - 2 T^a S_a + 2 V^a C_a = 0 \,,
\end{split}
\end{equation}
where we encounter the vector valued Cotton 2-form  defined as 
\begin{equation}\label{Cotton}
\begin{split}
C^a \equiv \mathcal{D} S^a \quad \longrightarrow \quad C_{a|bc} V^b V^c & =-\mathcal{D}_c S_{ab} V^b V^c + S_{al} {T^l}_{bc} V^b V^c \\
& = -\mathcal{D}_c S_{ab} V^b V^c - \frac{1}{6} S_{al} {\epsilon^{l}}_{bcd} \tilde{t}^d V^b V^c \,.
\end{split}
\end{equation}
In the case in which the torsion vanishes we recover the properties that the Schouten tensor is symmetric and that the completely antisymmetric and trace parts of the Cotton tensor are zero.

As a final check of the consistency of the theory
we can check from the previous equations that the Bianchi identities are identically satisfied. Indeed analyzing the torsion Bianchi in \eqref{Bianchifinal} by taking into account also \eqref{Gabfinal} together with \eqref{eqsfortorsfromconfinv1} and \eqref{eqsfortorsfromconfinv2} (recall that \eqref{Gabfinal}, \eqref{eqsfortorsfromconfinv1}, and \eqref{eqsfortorsfromconfinv2} come from conformal invariance of the theory, \emph{not} from the analysis of the field equations), we get $0=0$. One can prove, with some algebraic manipulation and making use of \eqref{proptors}, coming from the requirement of conformal invariance, that the same happens for the Bianchi of $G$. The Bianchi for $\tilde{T}^a$ becomes the Cotton Bianchi (for a connection with torsion) since $\tilde{T}^a$ coincides with the Cotton 2-form, while the Bianchi identity for $R^{ab}$ is the Bianchi for the Weyl 2-form $W^{ab}$ and simply leads to $\mathcal{R}^{ab} = 0$, the left-hand side of the latter being identically zero for any connection (here with a non-vanishing torsion).\footnote{The Weyl Bianchi written in the form $\mathcal{D} \mathcal{R}^{ab} = 0$ and the Cotton Bianchi look formally the same as in the case of vanishing torsion (see for instance \cite{Garcia:2003bw}).}

One can thus see that the Bianchi identities are, as expected, true identities, and they do not add any additional constraint to the theory. This result \eqref{Bianchifinal} was expected also from the vacuum analysis we have previously done.


\subsection{Equations of motion} 

Since we are adopting the second order formalism for $\omega$, that is $\omega=\omega(V,\mathfrak{D},\tilde{t})$, in particular by fixing the form of the torsion, we shall vary the Lagrangian \eqref{weyllagr} with respect to the independent fields $V$ and $\tilde{t}$.\footnote{In principle the spin connection $\omega^{ab}$ has a dilaton dependence, but, as we have previously discussed, the dilaton field  $\mathfrak{D}$ can be eliminated from the theory. Actually, as one can easily verify, even if $\mathfrak{D}$ is allowed to appear into the Lagrangian, exploiting the explicit form for the variation of $\omega^{ab}$ \eqref{deltaom} and the fact that the Weyl tensor is traceless, its dynamics is trivial, as expected.}

As we will see in a while, in this setup we get a propagation equation of the graviton that has four derivatives ($\partial^4 V$), as in the case with vanishing torsion, which is indeed expected for a conformal gravity theory. 

Now, let us observe that the following applies to our analysis:
\begin{equation}
\delta_{\Phi_i} \mathcal{L} = \frac{\delta \mathcal{L}}{\delta \omega} \frac{\delta \omega}{\delta \Phi_i} \delta \Phi_i + \frac{\delta \mathcal{L}}{\delta \Phi_i} \delta \Phi_i \,, \quad i=1,2 \,, \quad \Phi_1 = \tilde{t} \,, \Phi_2 = V \,, 
\end{equation}
schematically. The term $\frac{\delta \mathcal{L}}{\delta \Phi_i} \delta \Phi_i$ corresponds to the explicit variation of the various fields in the Lagrangian. Therefore, we may start by computing $\delta_\omega\mathcal{L}=\frac{\delta \mathcal{L}}{\delta \omega} \delta \omega$, where $\mathcal{L}$ is the torsionful Weyl Lagrangian \eqref{weyllagr}. Recalling eq. \eqref{RW} and using
\begin{equation}\label{deltaomegapart}
\begin{split}
\delta_\omega \mathcal{R}^{ab}&=\mathcal{D}(\delta\omega^{ab})= \mathcal{D}_c(\delta\omega^{ab})_dV^c V^d \,, \\
\delta_\omega \left({{\mathcal{R}}^{ab}}_{cd} \delta_b^d\right)&=\mathcal{D}_c (\delta \omega^{ab})_b \,, \\
\delta_\omega \left({{\mathcal{R}}^{ab}}_{cd} \delta_b^d \delta_a^c\right)&=\mathcal{D}_a ( \delta \omega^{ab})_b \,,
\end{split}
\end{equation}
 after partial integration and using the fact that the Weyl tensor is completely traceless, we find

\begin{equation}\label{deltaLom}
\delta_\omega\mathcal{L} = 8 \delta \omega_{ab|m} \left( \mathcal{D}_l W^{ablm} \right) \Omega^{(4)} \,,
\end{equation}
where we have also used the fact that
\begin{equation}
{\mathcal{D}}_b \Omega^{(4)} V^b = \frac{2}{3!} \epsilon_{defg} {T^d}_{bk} V^b V^k V^e V^f V^g = - \frac{1}{3 \cdot 3!} \epsilon_{dbkq} \tilde{t}^q \epsilon_{defg} \epsilon^{kefg} V^b \Omega^{(4)} = 0 \,,
\end{equation}
that is
\begin{equation}\label{DOmzero}
    {\mathcal{D}}_b \Omega^{(4)} = 0 \,.
\end{equation}

Regarding the variation with respect to $\tilde{t}^c$, one finds that the explicit variation does not contribute. Hence, from implicit variation one obtains 
\begin{equation}
\frac{\delta \mathcal{L}}{\delta \omega} \frac{\delta \omega}{\delta \tilde{t}^c} \delta \tilde{t^c} = - \frac{4}{3} \epsilon_{abmc} \delta \tilde{t}^c \left( \mathcal{D}_l W^{ablm} \right) \Omega^{(4)} = 0 \,,
\end{equation} 
which vanishes identically since  antisymmetrization of $W_{abcd}$ on three indices gives identically zero.\footnote{The symmetry properties of the Weyl tensor read
\begin{equation}\nonumber
\begin{split}
& W^{abcd} = - W^{abdc} = - W^{bacd} \,, \\
& W^{abcd} = W^{cdab} \,, \quad W^{a[bcd]} = 0 \,, \quad W^{[abcd]} = 0 \,.
\end{split}
\end{equation}
Note that they hold true also in the presence of torsion, since the Weyl tensor is an irrep of $\rm{SO}(1,3)$.}

Concerning the explicit variation with respect to $V$, here one can prove that we have
\begin{equation}\label{deltaVexpl}
\begin{split}
\frac{\delta \mathcal{L}}{(\delta V^p)_l} (\delta V^p)_l & = - 4 ( \delta V^p )_l \left( 2 \check{\mathcal{R}}_{ac} \delta^c_m \eta_{pb} - 2 \check{\mathcal{R}}_{ap} \eta_{bm} - \frac{2}{3} \mathcal{R} \eta_{am} \eta_{bp} \right) W^{ablm} \Omega^{(4)} \\
& = - 8 ( \delta V^p )_l \check{\mathcal{R}}_{ac} W^{aplc} \Omega^{(4)} \,,
\end{split}
\end{equation}
where we have also observed that the second and third term inside the round brackets in the first line give a vanishing contribution since they imply a tracing of the Weyl tensor. Then, using \eqref{deltaom}, \eqref{deltaLom} (and performing a partial integration), and \eqref{deltaVexpl}, together with the symmetry properties of the Weyl tensor and eq. \eqref{DOmzero}, after some algebraic manipulation we find that the field equation $\delta_{V} \mathcal{L}=0$ reads
\begin{equation}\label{deltaVcomplete}
( \delta V^p )_l \left( \mathcal{D}_q \mathcal{D}_t W^{tpql} - \frac{1}{2} \check{\mathcal{R}}_{ac} W^{aplc} - \frac{1}{12} \epsilon^{abpc} \tilde{t}_c \mathcal{D}_t {W_{ab}}^{tl} \right) = 0 \,,
\end{equation}
namely
\begin{equation}\label{propV}
\mathcal{D}_q \mathcal{D}_t W^{tpql} - \frac{1}{2} \check{\mathcal{R}}_{ac} W^{aplc} - \frac{1}{12} \epsilon^{abpc} \tilde{t}_c \mathcal{D}_t {W_{ab}}^{tl} = 0 \,.
\end{equation}
Notice that the $p,l$ trace of the latter identically vanishes, due to the symmetry properties of the Weyl tensor.
The first two terms in \eqref{propV} are \emph{formally} the same as in the absence of torsion, thus giving a fourth order equation for the vierbein. At linearized level the kinetic term is actually the same as in the absence of torsion, while at higher level the presence of contorsion in the spin connection gives higher order corrections.

Hence, in our theory we get a fourth order equation for the vierbein and in the limit in which $\tilde{t}_a$ is set to zero we recover the conformal theory of \cite{Kaku:1977pa}.\footnote{Let us observe that, as we have previously mentioned, the same eqs. \eqref{proptors} and \eqref{propV} can be obtained by gauging \eqref{vacuumnew} and implementing the constraints we have presented in our analysis in order to recover conformal invariance.}

\section{On the extension to conformal supergravity}\label{prelres}

In this section we give some preliminary results concerning conformal $\mathcal{N}=1$, $D=4$ supergravity in the geometric approach. As we have already mentioned, in the literature, besides vanishing supertorsion, some constraints have been implemented to recover superconformal invariance (the same constraints have been implemented in \cite{Castellani:1981um}, within the geometric approach, through Lagrange multipliers).
 
Here we start a preliminary analysis at the level of Bianchi identities using the geometric approach, showing that all the aforementioned constrained can be directly obtained from the study of the Bianchi identities, just imposing the vanishing of the supertorsion (which can be viewed as the direct supersymmetric extension of the constraint of vanishing torsion imposed in \cite{Kaku:1977pa}).\footnote{For the original formulation of the geometric approach to supergravity in superspace and, in particular, of its application to the study of the Bianchi identities in superspace we refer the reader to \cite{Castellani:1991et,DAuria:2020guc} (see also \cite{Castellani:1981um} and Appendix A and B of \cite{Andrianopoli:1996cm}). Moreover, a concise review of the prescriptions on the supercurvatures in the geometric approach to supergravity is also given in Appendix A of Ref. \cite{Andrianopoli:2014aqa}.} 

For the benefit of the reader and in order to establish our formalism let us just recall the main basic points of the geometric approach to supergravity.

The gauge fields are now super 1-forms in superspace that can be expanded along the supervierbein $(V^a,\psi^\alpha)$, with $\alpha=1,\ldots ,4$ and $\psi^\alpha$ being the gravitino 1-form\footnote{Here we are considering $\mathcal{N}=1$, $D=4$. Spinor indices are denoted by $\alpha,\beta, \ldots$ and in the sequel we will frequently omit them to lighten the notation.} (note that in the geometric approach the superfields are never expanded in terms of the Grassmann coordinates).
Analogously, the supercurvatures  are 2-forms which can be expanded along basis of 2-forms, namely
\begin{equation}\label{genexpcurv}
R^A = {R^A}_{ab} V^a V^b + {R^A}_{a\alpha} V^a \psi^\alpha + {R^A}_{\alpha \beta} \psi^\alpha \psi^\beta \,,
\end{equation}
where ${R^A}_{a\alpha}$ and ${R^A}_{\alpha \beta}$ are the outer components of $R^A$, while ${R^A}_{ab}$ are the inner ones.\footnote{The outer components of the curvatures are defined as those having at least one index along the $\psi$ direction of superspace, while the components with indices only along the bosonic vierbein are called inner.}
The important point is that, both in the Lagrangian approach as well as in the Bianchi identities approach, it turns out that all the outer components of the curvatures can be expressed algebraically in terms of the inner ones, thus allowing for the elimination of the spurious unphysical degrees of freedom from the theory.\footnote{The relation between outer and inner comoponents of the supercurvatures is also referred to as the ``rheonomy principle''. Actually this property is a consequence of the fact that the Lagrangian is constructed only in terms of differential 4-forms in superspace, with the exclusion of the Hodge duality operator.}
Actually, this can be shown from both the study of the geometric Lagrangian and the sector-by-sector analysis of the Bianchi ``identities''. Within the latter approach, the Bianchi identities become relations to be analyzed performing their split in the different sectors $\psi \psi \psi$, $\psi \psi V$, $\psi V V$, and $V V V$. This gives the expression of the outer components of the supercurvatures in terms of the inner ones, making the theory on superspace have the same physical content as the theory on spacetime. 

Finally, we mention that since supersymmetry transformations are just Lie derivatives in superspace, they are easily derived from the (superspace) Lie derivative of the gauge fields using the formula in footnote 7, namely $\delta_\epsilon \mu^A = \mathrm{D} \epsilon^A + \imath_\epsilon R^A$, where $ \mathrm{D} \epsilon^A $ is a gauge transformation and the contraction is made with a supersymmetry parameter.

We shall now apply the aforementioned prescription on the Bianchi identities to the case of conformal supergravity with vanishing supertorsion.

\subsection{Bianchi identities of the superconformal group with vanishing supertorsion}

The superconformal algebra \cite{Wess:1974tw,Ferrara:1974qk} is generated by the set $\lbrace{ \mathbf{J}_{ab}, \mathbf{P}_a, \mathbf{K}_a, \mathbf{D}, \mathbf{A}, \mathrm{Q}_\alpha, \mathcal{Q}_\beta \rbrace} $. 
We introduce the 1-form fields $\lbrace{\omega^{ab}, V^a, K^a, \mathfrak{D}, A, \psi^\alpha , \phi^\alpha\rbrace}$ (see also \cite{Kaku:1978nz,Townsend:1979ki,Castellani:1981um}), respectively dual to the vector fields generators of the superconformal algebra as given by \eqref{dualgen} together with
\begin{equation}\label{dualgensuper}
A(\mathbf{A}) = 1 \,, \quad \psi^\alpha({\mathrm{Q}}_\beta)=\delta^\alpha_\beta \,, \quad \phi^\alpha(\mathcal{Q}_\beta)=\delta^\alpha_\beta \,.
\end{equation}
The scale weight of the $\rm{U}(1)$ gauge 1-form field $A$, of the gravitino 1-form $\psi$, and of the \emph{conformino} 1-form $\phi$ are, respectively,
\begin{equation}\label{formscalesuper}
[A]=0 \,, \quad [\psi]= \frac{1}{2} \,, \quad [\phi]= - \frac{1}{2} \,.
\end{equation}
The supercurvatures associated with the superconformal algebra are
\begin{equation}\label{supercurvatures}
\begin{split}
R^{ab} & \equiv \mathcal{R}^{ab} - 4 V^{[a} K^{b]} + \bar{\psi} \gamma^{ab} \phi \,, \\
T^a & \equiv \mathcal{D} V^a + \mathfrak{D} \wedge V^a - \frac{\ii}{2} \bar{\psi} \gamma^a \psi = \hat{\mathcal{D}} V^a - \frac{\ii}{2} \bar{\psi} \gamma^a \psi \,, \\
\mathcal{T}^a & \equiv \mathcal{D} K^a - \mathfrak{D} \wedge K^a + \frac{\ii}{2} \bar{\phi} \gamma^a \phi = \hat{\mathcal{D}} K^a + \frac{\ii}{2} \bar{\phi} \gamma^a \phi \,, \\
G & \equiv d \mathfrak{D} + 2 V^a K_a - \bar{\psi}\phi \,, \\
F & \equiv dA + 2 \ii \bar{\psi} \gamma^5 \phi \,, \\
\rho & \equiv \mathcal{D} \psi + \frac{1}{2} \mathfrak{D} \wedge \psi - \frac{3 \ii}{4} A \gamma^5 \psi - \ii \gamma_a \phi V^a = \hat{\mathcal{D}} \psi - \frac{3 \ii}{4} A \gamma^5 \psi - \ii \gamma_a \phi V^a = \nabla \psi - \ii \gamma_a \phi V^a \,, \\
\sigma & \equiv \mathcal{D} \phi - \frac{1}{2} \mathfrak{D} \wedge \phi + \frac{3 \ii}{4} A \gamma^5 \phi + \ii \gamma_a \psi K^a = \hat{\mathcal{D}} \phi + \frac{3 \ii}{4} A \gamma^5 \phi + \ii \gamma_a \psi K^a = \nabla \phi + \ii \gamma_a \psi K^a \,,
\end{split}
\end{equation}
where $\psi^\alpha$ and $\phi^\alpha$ are the gravitino and conformino 1-forms, dual to ordinary supersymmetry and conformal supersymmetry, respectively. We recall that $\mathcal{D}=d-\omega$ is the Lorentz covariant derivative, $\hat{\mathcal{D}}$ is the Lorentz plus scale covariant derivative, and  we have also taken the opportunity to introduce, besides, a Lorentz plus scale plus $\rm{U}(1)$ covariant derivative $\nabla$. The matrices $\gamma^a$, $\gamma^{ab}$, and $\gamma^5$ are the usual gamma matrices in four dimensions. Useful formulas on gamma matrices can be found in Appendix \ref{gammamatr}.  

The Bianchi identities obeyed by the supercurvatures \eqref{supercurvatures} are
\begin{equation}\label{Bianchisuper}
\begin{split}
& \mathcal D R^{ab} +4 (T^{[a} K^{b]} -V^{[a} \mathcal T^{b]}) +\bar \phi\gamma^{ab}\rho +\bar \psi\gamma^{ab}\sigma =0 \,, \\
& \hat{\mathcal{D}} T^a +  {R}^{ab}V_b - G V^a -\ii \bar\psi\gamma^{a}\rho=0 \,, \\
& \hat{\mathcal{D}} {\mathcal T^a} +  {R}^{ab}K_b + G K^a +\ii \bar\phi\gamma^{a}\sigma=0 \,, \\
& d G -2 T^a K_a + 2 V^a\mathcal T_a -\bar\psi \sigma +\bar\phi \rho=0 \,, \\
& d F + 2 \rm{i}\bar\psi \gamma^5\sigma - 2 \rm{i}\bar\phi \gamma^5\rho=0 \,, \\
& \nabla \rho +\frac{1}{4}\gamma_{ab}{R}^{ab} \psi -\frac12 G\psi +\frac{3\rm{i}}{4} F\gamma^5\psi +{\rm{i}} \gamma_a \sigma V^a - {\rm{i}} \gamma _a \phi T^a=0 \,, \\
& \nabla \sigma +\frac{1}{4}\gamma_{ab}{R}^{ab} \phi +\frac12 G\phi -\frac{3\rm{i}}{4} F\gamma^5\phi -{\rm{i}} \gamma_a \rho K^a + {\rm{i}} \gamma _a \psi \mathcal{T}^{a}=0 \,,
\end{split}
\end{equation}
where
\begin{equation}
\begin{split}
& \hat{\mathcal{D}} T^a \equiv \mathcal{D} T^a + \mathfrak{D} \wedge T^a \,, \\
& \hat{\mathcal{D}} \mathcal{T}^a \equiv \mathcal{D} \mathcal{T}^a - \mathfrak{D} \wedge \mathcal{T}^a \,, \\
& \nabla \rho \equiv \mathcal{D} \rho + \frac{1}{2} \mathfrak{D} \wedge \rho - \frac{3\ii}{4} A \gamma^5 \rho = d \rho - \frac{1}{4} \gamma_{ab} R^{ab} \rho + \frac{1}{2} \mathfrak{D} \wedge \rho - \frac{3\ii}{4} A \gamma^5 \rho \,, \\
& \nabla \sigma \equiv \mathcal{D} \sigma - \frac{1}{2} \mathfrak{D} \wedge \sigma + \frac{3\ii}{4} A \gamma^5 \sigma = d \sigma - \frac{1}{4} \gamma_{ab} R^{ab} \sigma - \frac{1}{2} \mathfrak{D} \wedge \sigma + \frac{3\ii}{4} A \gamma^5 \sigma \,.
\end{split}
\end{equation}
One can now apply the prescription on the Bianchi identities to the present case, that is writing the supercurvatures expansion as given in \eqref{genexpcurv} and differentiating it to compare the result with the Bianchi \eqref{Bianchisuper} expanded along the supervierbein basis. The closure of the resulting system of equations must occur sector-by-sector, that is along the $\psi \psi \psi$, $\psi \psi V$, $\psi V V$, and $V V V$ sectors separately.

Imposing vanishing supertorsion ($T^a=0$) from the very beginning, a careful analysis shows that the superspace curvatures must have the following parametrization:\footnote{The scale of the supercurvatures components along the 2-vierbein sector is 
\begin{equation}\nonumber
[{R^{ab}}_{cd}, G_{ab}, F_{ab}]=-2\,,\quad [{T^a}_{bc}]=-1\,,\quad [{\mathcal{T}^a}_{bc}]=-3\,,\quad [\rho_{ab}]=-\frac{3}{2}\,,\quad [\sigma_{ab}]=-\frac{5}{2} \,.
\end{equation}
When doing the explicit calculations one can immediately simplify the starting general Ansatz by exploiting scale weight arguments.}
\begin{equation}\label{parvanishingtor}
\begin{split}
R^{ab} & = {R^{ab}}_{cd} V^c V^d + 2 \ii \bar\psi \gamma_c \rho^{ab} V^c  \,, \\
T^a & = 0 \,, \\
\mathcal T^a & = {\mathcal{T}^a}_{bc}V^b V^c + \bar\psi \left( - \frac{1}{2} \sigma^{ab} - \frac{\ii}{2} \gamma^5 \, {}^\star{\sigma}^{ab} + \gamma^{(a} \gamma_m \sigma^{m|b)} \right) V_b \,, \\
G & = G_{ab} V^a V^b \,, \\
F & = F_{ab} V^a V^b \,, \\
\rho & = \rho_{ab} V^a V^b \,, \\
\sigma & = \sigma_{ab}V^a V^b + \left( - \frac{\ii}{2} {}^\star{F}_{ab} \gamma^b + \frac{1}{2} F_{ab} \gamma^5 \gamma^b \right) \psi V^a \,,
\end{split} 
\end{equation}
where for any 0-form $U_{ab}=-U_{ba}$ we have denoted the corresponding Hodge-dual as
\begin{equation}
{}^\star{U}_{ab} = \frac{1}{2} \epsilon_{abcd} U^{cd} \,.
\end{equation}
As previously observed, the supersymmetry transformation laws differ from the gauge transformations when the curvatures exhibit at least a gravitino $\psi$ in their parametrization. In particular, in the case at hand this happens for $R^{ab}$, $\mathcal{T}^a$, and $\sigma$, which indeed have a $\psi$ in their parametrization (for the explicit form of the supersymmetry transformations of the fields we refer the reader to \cite{Kaku:1978nz,Castellani:1981um}).

Let us recall here that the quantities ${R^{ab}}_{cd}$, ${\mathcal{T}^a}_{bc}$, $G_{ab}$, $F_{ab}$, $\rho_{ab}$, and $\sigma_{ab}$ appearing in the parametrization \eqref{parvanishingtor} are the so-called \emph{supercovariant field-strengths} and they differ, in general, from the spacetime projections of the supercurvatures. Indeed, let us refer e.g. to the Lorentz supercurvature. Taking the components of $R^{ab}$ along $dx^\mu\wedge dx^\nu$, namely ${R^{ab}}_{\mu \nu}={R^{ab}}_{cd} V^c_\mu V^d_\nu + 2 \ii \bar{\psi}_{[\mu} \gamma_c \rho^{ab} V^c_{\nu]} $, we see that the spacetime components ${R^{ab}}_{\mu\nu}$ differ from the components along the purely bosonic supervierbein, ${R^{ab}}_{cd} V^c_\mu V^d_\nu$. The quantity ${R^{ab}}_{cd} V^c_\mu V^d_\nu\equiv{R^{ab}}_{\mu \nu|(\text{cov})}={R^{ab}}_{cd} V^c_\mu V^d_\nu + 2 \ii \bar{\psi}_{[\mu}\gamma_c\rho^{ab}V^c_{\nu]}$ is the supercovariant field-strength. The same happens for the curvatures $\mathcal T^a$ and $\sigma$. Instead, as in the present case the parametrizations of $G$, $F$, and $\rho$ have just components along two vierbein , covariant and supercovariant components on spacetime are identified, that is we have $G_{\mu \nu}=G_{ab} V^a_\mu V^b_\nu$, $F_{\mu \nu}=F_{ab} V^a_\mu V^b_\nu$, and $\rho_{\mu \nu} = \rho_{ab} V^a_\mu V^b_\nu$.

Besides the given parametrizations one also obtains the following constraints:
\begin{equation}\label{GFR}
G_{ab} = \frac{1}{2} {}^\star{F}_{ab} \,, \quad \check{R}_{[ab]} = - G_{ab} = - \frac{1}{2} {}^\star{F}_{ab} \,, \quad \check{R}_{(ab)} = 0 \,, \quad R = 0 \,,
\end{equation}
\begin{equation}\label{sigmaconstr}
\begin{split}
& \gamma^{ab} \sigma_{ab} = 0 \,, \\
& \gamma^a \left( \sigma_{ab} - \ii \gamma^5 {}^\star{\sigma}_{ab} \right) = 0 \,, \\
& \gamma_{m[a} \sigma_{m|b]} = \frac{1}{2} \left( \sigma_{ab} - \ii \gamma^5 {}^\star{\sigma}_{ab} \right) \,,
\end{split}
\end{equation}
and
\begin{equation}
\gamma_{[a} \rho_{bc]} = 0 \quad \longrightarrow \quad \gamma_c \rho_{ab} = - 2 \gamma_{[a} \rho_{b]c} \,,    
\end{equation}
the latter implying
\begin{equation}
\begin{split}\label{rhoconstr}
& \gamma^a \rho_{ab} = 0 \quad \Rightarrow \quad \gamma^{ab} \rho_{ab} = 0 \,,  \\
& \rho_{ab} + \ii \gamma^5 {}^\star{\rho}_{ab} = 0 \,. \end{split}
\end{equation}
Notice that using the first equation of \eqref{sigmaconstr}, after some algebraic manipulation, we get that the second of \eqref{sigmaconstr} reduces to the trivial identity $0=0$. 

Let us just give a brief summary of the main steps of the cumbersome calculations to recover the above results. The parametrization of $R^{ab}$, $G$, and $F$, together with the constraints in \eqref{rhoconstr} and the fact that $\sigma$ does not have components along two $\psi$, can be obtained by analyzing the $\psi \psi \psi$ sector of the Bianchi for $R^{ab}$, $G$, and $F$ together with the $\psi \psi V$ sector of the Bianchi for $T^a$ and $\rho$. Considering the $\psi \psi V$ sector of the Bianchi identities for $G$ and $F$ together with the $\psi \psi \psi$ sector of the Bianchi for $\sigma$, the $\psi V V $ sector of the Bianchi for $\rho$, the $\psi \psi V$ sector of the Bianchi for $R^{ab}$, and the $VVV$ sector of the supertorsion Bianchi, one finds \eqref{GFR} and the parametrization of $\sigma$. Finally, the parametrization of $\mathcal{T}^a$ and the equations in \eqref{sigmaconstr} can be obtained by analyzing the $\psi V V$ sector of the Bianchi for $G$ and $F$ together with the $\psi \psi V$ sector of the Bianchi for $\sigma$.

The above results are in perfect agreement with the ones of \cite{Kaku:1978nz,Townsend:1979ki}  and \cite{Castellani:1981um}. In particular, the constraints in \eqref{rhoconstr} are the ones used in \cite{Kaku:1978nz,Townsend:1979ki} (together with $T^a=0$). Moreover, the constraint $\gamma^a \rho_{ab} = 0$ in \eqref{rhoconstr}, is the same constraint fixed in \cite{Castellani:1981um} by using Lagrange multipliers in the Lagrangian in order to recover superconformal invariance of the theory.
We conclude that the solution given by eqs. \eqref{parvanishingtor}, \eqref{GFR}, \eqref{sigmaconstr}, and \eqref{rhoconstr} gives exactly the same results as in Refs. \cite{Kaku:1978nz,Townsend:1979ki}  and \cite{Castellani:1981um}. There the constraints were required by physical arguments in order to have consistently supersymmetry invariance, while here we have shown that they are a mere consequence of the geometrical structure of the theory expressed by the closure of the Bianchi identities.

Observe that the constraints derived from the Bianchi identities turn out to be necessary for their closure, in a way quite analogous to the requirement that in the absence of auxiliary fields the closure of the supergravity Bianchi identities only holds when the equations of motion are satisfied. However, in conformal supergravity the parametrizations for the curvatures and the constraints recovered so far do not imply the equations of motion. One could then be surprised that we need constraints to have closure, since, after all, Bianchi identities, when no equation of motion is needed, are true identities. The point is that the Bianchi identities would be true identities if we analyzed them in the enlarged \emph{superconformal} coset of the basis gauge fields $( V^a, K^a,\psi^\alpha,\phi^\alpha )$, the other gauge fields $\omega^{ab}$, $\mathfrak{D}$, and $A$ being the factorized 1-forms dual to the generators belonging to the fiber. However, we want to have a physical theory on the \emph{ordinary} supercoset spanned only by $(V^a,\psi^\alpha)$, which is a \emph{cotangent submanifold} of the enlarged superconformal coset. The geometric constraints of theory are then interpreted as the requirement needed in order to have a consistent projection from the superconformal coset into the ordinary superspace.

The fact that the study of the Bianchi identities leads to the constraint of conformal supergravity has been also inferred in \cite{Cribiori:2018xdy} in the context of an off-shell formulation of $N=2$ supergravity with tensor multiplets. 
Here we have further highlighted and clarified the geometric origin and meaning of the superconformal constraints in the case of $\mathcal{N}=1$, $D=4$ conformal supergravity with $T^a=0$, whose understanding is rather fundamental in view of a future analysis including a non-vanishing supertorsion in the theory.

Let us also mention that there are no independent differentials in the $\mathbf{K}_a$ and $\mathcal{Q}_\alpha$ directions (as can be also deduced looking at \eqref{supercurvatures} and \eqref{Bianchisuper}) so that one can write,  using also scale weight arguments,

\begin{equation}\label{Kaphipar}
\begin{split}
K^a & = {K^a}_b V^b + \bar{\psi} \kappa^a \,, \\     
\phi & = \phi_a V^a \,,    
\end{split}
\end{equation}
where the 0-forms ${K^a}_b $, $\kappa^a$, and $\phi_a$ are a tensor, a spinor vector, and another spinor vector, respectively. Recall that ${K^a}_b $ coincides with the spacetime components of $K^a$ only when $\psi \rightarrow 0$, but since we are now in superspace the whole spacetime components of $K^a$ is given by the supercovariant part of $K^a$ (that is ${K^a}_b$ in \eqref{Kaphipar}) plus the component along $\psi$. When one formulates the Lagrangian for the theory, the above components of $K^a$ and $\phi$ can be determined by studying the field equations of the theory (and this could be particularly useful in our future study where we will try to include a non-vanishing supertorsion).

On the other hand, the aforementioned components can be also obtained by expanding the supercurvature definitions given in \eqref{supercurvatures} and using the geometric parametrization \eqref{parvanishingtor}. For the conformino components $\phi_a$ we get
\begin{equation}\label{phibvaninshingtor}
\phi_b = \frac{2}{3} \gamma^a \left( \ii \rho_{(0)|ab} - \frac{1}{2} \gamma^5 \, {}^\star{\rho}_{(0)|ab} \right) \,,
\end{equation}
where we have exploited $\gamma^a \rho_{ab}=0$ from \eqref{rhoconstr} and used the definition of $\rho$ given in \eqref{supercurvatures} taking its 2-vierbein sector (which is the only sector appearing into the parametrization of $\rho$ in \eqref{parvanishingtor}), that is\footnote{We note that since there are no components of $\rho$ along the outer basis $(\psi , V)$ and $(\psi , \psi)$, we have $\rho_{\mu\nu}=\rho_{ab} V^a_\mu V^b_\nu$ and therefore we can identify the $a,b$ indices with spacetime anholonomic indices related to each other by the four-dimensional vierbein. This observation explains the meaning of the subsequent equation $\rho_{(0)|ab} \equiv \nabla_{[a} \psi_{b]}$ in \eqref{rho2v}, which would be senseless if $a,b$ were interpreted as superspace indices along $V^a V^b$ since the 1-form $\psi$ in superspace is independent of $V^a$ by definition.}
\begin{equation}\label{rho2v}
\rho_{ab} = \rho_{(0)|ab} + \ii \gamma_{[a} \phi_{b]}  \,, \quad \rho_{(0)|ab} \equiv \nabla_{[a} \psi_{b]} \,.
\end{equation}
Eq. \eqref{phibvaninshingtor} coincides, up to normalization and conventions, with the expression for the conformino ($\phi_\mu=\phi_b V^b_\mu$) found in \cite{Kaku:1978nz}.
Finally, notice that using the other results on $\rho_{ab}$ given in \eqref{rhoconstr}, making some algebraic manipulation we also obtain
\begin{equation}
\rho_{(0)|ab} + \ii \gamma^5 \, {}^\star{\rho}_{(0)|ab} = 0 \,. \label{rho0constr}
\end{equation}
Thus, using \eqref{rho0constr} into \eqref{phibvaninshingtor}, we are left with
\begin{equation}\label{phibvaninshingtornew}
\phi_b = \frac{\ii}{3} \gamma^a \rho_{(0)|ab} \,.
\end{equation}
Similar arguments can be applied to find the expression for $K^a={K^a}_\mu d x^\mu ={K^a}_b V^b + \bar{\psi} \kappa^a $ by looking at the definition of $R^{ab}$ in \eqref{supercurvatures} and using the parametrization for $R^{ab}$ in \eqref{parvanishingtor}. More precisely, defining
\begin{equation}\label{R0def}
{\mathcal{R}}^{ab}_{(0)} \equiv \mathcal{R}^{ab} + \bar{\psi} \gamma^{ab} \phi \,,    
\end{equation}
in such a way that fermionic contributions are taken into account in a straightforward way by means of ${\mathcal{R}}^{ab}_{(0)}$, we get
\begin{equation}\label{Kabsuper}
\begin{split}
& \quad {R^{ab}}_{\mu \nu} - {\mathcal{R}^{ab}}_{(0)|\mu \nu} = - 4 {V^{[a}}_{[\mu} {K^{b]}}_{\nu]} \\
\quad \Rightarrow & \quad K_{\mu \nu} = K_{a\mu} V^a_\nu = \frac{1}{2} \left(\check{\mathcal{R}}_{(0)|\mu \nu} - \frac{1}{6} g_{\mu \nu} {\mathcal{R}}_{(0)} \right) - \frac{1}{2} \check{R}_{[\mu \nu]} - \frac{\ii}{2} \bar{\psi}^\lambda \gamma_\nu \rho_{\lambda \mu} \\
\quad \phantom{\rightarrow} & \quad \phantom{K_{\mu \nu} = K_{a\mu} V^a_\nu} = \frac{1}{2} \left(\check{\mathcal{R}}_{(0)|\mu \nu} - \frac{1}{6} g_{\mu \nu} {\mathcal{R}}_{(0)} \right) + \frac{1}{4} {}^\star{F}_{\mu \nu} - \frac{\ii}{2} \bar{\psi}^\lambda \gamma_\nu \rho_{\lambda \mu} \,.
\end{split}
\end{equation}
The latter coincides, up to normalization and conventions, with the same expression found in \cite{Kaku:1978nz} for $K_{\mu \nu}$.

We have thus shown that at the supersymmetric level, setting the supertorsion to zero, all the other constraints necessary for superconformal invariance and implemented in \cite{Castellani:1981um} through Lagrange multipliers here actually follows geometrically, from the study of the various sectors of the Bianchi identities. Therefore, we expect the Bianchi identities to be a key feature in order to explore the possible construction of a conformal supergravity theory with a non-vanishing supertorsion, dictating, in this framework, the constraints that one must impose on the theory in order to recover superconformal invariance.

\section{Conclusions}\label{disc}

In this paper we have shown that, in contrast with the claim of Ref. \cite{Kaku:1977pa}, it is actually possible to construct a gauge theory of the conformal group in four spacetime dimensions with a non-vanishing torsion component. In particular, we have allowed for a non-vanishing axial vector torsion and found a sufficient condition to write a gauge theory for the conformal group. In this setup, invariance under proper special conformal transformations (conformal boosts) implies a Killing equation for the axial vector torsion that, upon further differentiation, leads to a Maxwell-like propagation equation (precisely, to the d'Alembert equation \eqref{proptors}) in a curved background for the aforementioned axial vector.
In the limit in which $\tilde{t}_a$ is set to zero we recover the conformal theory of \cite{Kaku:1977pa}. 


We have then given some preliminary results regarding conformal $\mathcal{N}=1$, $D=4$ supergravity. In particular, we have shown that the constraints introduced in \cite{Castellani:1981um} in the geometric approach by the use of Lagrange multipliers in order to recover superconformal invariance of the theory can be, in fact, directly obtained in the same geometric approach from the study of the Bianchi identities, just assuming vanishing supertorsion.

This paves the way for future investigations that will be devoted to deepening the analysis on conformal supergravity (this work is currently in progress \cite{RL}). In particular, since we have seen that at the purely bosonic level there exist the possibility of introducing a non-vanishing completely antisymmetric torsion without spoiling conformal invariance, we argue that something similar may also occur in the superconformal case. 
As a further remark, from a first glance we can say that (some of) the supersymmetric constraints arising from the requirement of superconformal invariance would certainly be different from the ones obtained in the case in which the supertorsion is set to zero in order to get a superconformal theory, and this in particular might cause something unexpected to happen. 

Finally, let us say that our findings could also prove useful in the development a possible four-dimensional extension of the theories presented in \cite{Klemm:2020mfp,Klemm:2020gfm} in the context of modified/alternative theories of gravity.
A detailed study in this direction could also unveil some peculiar features of non-Riemannian degrees of freedom, together with a clearer understanding of the potential relations occurring at the dynamical level among them.


\section*{Acknowledgements}

We thank our friends and colleagues Laura Andrianopoli and Mario Trigiante for a critical reading of the manuscript, useful suggestions, and interesting discussions.
L.R. would like to thank the Department of Applied Science and Technology of the Polytechnic University of Turin, and in particular Laura Andrianopoli and Francesco Raffa, for financial support.


\appendix

\section{Useful formulas on gamma matrices}\label{gammamatr}

We are working with Majorana spinors, satisfying $ \bar \lambda = \lambda^T C$, where $C$ is the charge conjugation matrix.

\begin{itemize}
\item \textit{Symmetric gamma matrices}: $C \gamma_a$, $C\gamma_{ab}$, $C \gamma_5 \gamma_{ab}$.
\item \textit{Antisymmetric gamma matrices}: $ C $, $C\gamma_5$, $C\gamma_5 \gamma_{a}$.
\item \textit{Clifford algebra}:
\begin{equation}
\begin{split}
& \lbrace{\gamma_a,\gamma_b \rbrace} = 2\eta_{ab}  \,, \quad  \left[\gamma_a,\gamma_b \right] = 2 \gamma_{ab} \,, \quad \gamma_5 \equiv - \ii \gamma_0 \gamma_1 \gamma_2 \gamma_3  \,, \\
& \gamma _0^{\dagger} = \gamma _0 \,, \quad \gamma _0 \gamma _i^{\dagger}  \gamma _0 = \gamma _i \quad (i=1,2,3) \,, \quad \gamma _5^{\dagger} = \gamma _5 \,, \\
& \epsilon _{abcd} \gamma^{cd} = 2 \ii \gamma_{ab} \gamma_5 \,, \quad \gamma_{ab} \gamma_5 = \gamma_5 \gamma_{ab} \,, \quad \gamma_a \gamma_5 = - \gamma_5 \gamma_{a} \,, \\
& \gamma_m \gamma^{ab}\gamma^m =0 \,, \quad \gamma_{ab}\gamma_m \gamma^{ab}=0   \,, \quad      \gamma_{ab}\gamma_{cd} \gamma^{ab}= 4 \gamma_{cd} \,, \quad \gamma_m \gamma^a \gamma^m = -2\gamma^a \,, \\
& \gamma^a \gamma_a = 4 \,, \quad \gamma_b \gamma^{ab} = - 3 \gamma^a \,, \quad \gamma^{ab} \gamma_b = 3 \gamma^a \,, \\
& \gamma^{ab}\gamma^c =2 \gamma^{[a}\delta^{b]}_c + \ii \epsilon^{abcd}\gamma_5 \gamma_d \,, \quad \gamma^c \gamma^{ab}= -2 \gamma^{[a}\delta^{b]}_c + \ii \epsilon^{abcd}\gamma_5 \gamma_d \,, \\
& \gamma_{ab}\gamma_{cd} =  \ii \epsilon^{abcd}\gamma_5 -4\delta^{[a}_{[c}\gamma^{b]}_{\ d]} -2\delta^{ab}_{cd} \,.
\end{split}    
\end{equation}
\item \textit{Useful Fierz identities for $\mathcal{N}=1$} (for the 1-form spinor $\psi$):
\begin{equation}
\begin{split}
& \psi \bar\psi =  \frac 14 \gamma_a\bar\psi \gamma^a\psi-\frac 18 \gamma_{ab}\bar\psi\gamma^{ab}\psi \,, \\
& \gamma_a\psi\bar\psi \gamma^a\psi = 0 \,, \\
& \gamma_{ab}\psi\bar\psi\gamma^{ab}\psi = 0 \,.
\end{split}
\end{equation}
Irreducible 3-$\psi$ representations:
\begin{equation}
\begin{split}
& \Xi^{a}_{(12)} \equiv \psi\bar\psi \gamma^a\psi \,, \\
& \Xi^{ab}_{(8)} \equiv \psi\bar\psi \gamma^{ab}\psi + \gamma^{[a} \Xi^{b]}_{(12)} \,.
\end{split}    
\end{equation}
They satisfy $\gamma_a \Xi^{a}_{(12)}=0$, $\gamma_a \Xi^{ab}_{(8)}=0$, and we further have
\begin{equation}
\gamma_{ab}\psi\bar\psi \gamma^a\psi =  - \gamma^a\psi\bar\psi \gamma_{ab}\psi = - \gamma_5\gamma^a \psi\bar\psi \gamma_{ab}\gamma_5\psi= \Xi^{(12)}_b \,.
\end{equation}
\item \textit{Some spinor identities}:
\begin{equation}
\begin{split}
& \bar{\psi} \xi = \left(-1\right)^{pq} \bar{\xi} \psi \,, \\
& \bar{\psi} (S) \xi = - \left(-1\right)^{pq} \bar{\xi} (S) \psi \,, \\
& \bar{\psi} (AS) \xi = \left(-1\right)^{pq} \bar{\xi} (AS) \psi \,, 
\end{split}    
\end{equation}
where $(S)$ is a symmetric matrix, while $(AS)$ is an antisymmetric one; $\psi$ and $\xi$ are, respectively, a generic $p$-form spinor and a generic $q$-form spinor.
\end{itemize}


\end{document}